# Temperature Size Rule is mediated by thermal plasticity of critical size in *Drosophila melanogaster*

Shampa M. Ghosh[1*], Nicholas D. Testa[1], Alexander W. Shingleton[1]

[1]*Department of Zoology, Michigan State University, East Lansing, MI 48824, USA*

Emails: modak@msu.edu; testanic@msu.edu; shingle9@msu.edu

[*] For correspondence:
Shampa M. Ghosh
Ecology, Evolutionary Biology & Behavior
Department of Zoology
240 Natural Science
Michigan State University
East Lansing, MI 48824, USA
Phone: 517 353 5478
Fax: 517 432 2789
E-mail: modak@msu.edu

**Abstract**

Most ectotherms show an inverse relationship between developmental temperature and body size, a phenomenon known as the temperature size rule (TSR). Several competing hypotheses have been proposed to explain its occurrence. According to one set of views, the TSR results from inevitable biophysical effects of temperature on the rates of growth and differentiation, whereas other views suggest the TSR is an adaptation that can be achieved by a diversity of mechanisms in different taxa. Our data reveal that the fruit fly, *Drosophila melanogaster*, obeys the TSR using a novel mechanism: reduction of critical size at higher temperatures. In holometabolous insects, attainment of critical size initiates the hormonal cascade that terminates growth, and hence, *Drosophila* larvae appear to instigate the signal to stop growth at a smaller size at higher temperatures. This is in contrast to findings from another holometabolous insect, *Manduca sexta*, in which the TSR results from the effect of temperature on the rate and duration of growth. This contrast suggests that there is no single mechanism that accounts for the TSR. Instead, the TSR appears to be an adaptation that is achieved at a proximate level through different mechanisms in different taxa.

## Introduction

Body size is a major organismal trait that affects multiple aspects of an animal's biology, from its anatomy and physiology, to its behavior and ecology [1-3]. Body size also shows high levels of plasticity in response to the developmental environment [4-8]. One environmental factor that has a particularly dramatic effect on body size, at least in ectotherms, is temperature: in almost all ecotherms an increase in developmental temperature leads to a decrease in final adult size [9, 10]. The phenomenon is so general that it has been dubbed the *temperature size rule* (TSR) [9] although, as with all 'rules', there are some exceptions [11, 12]. Nevertheless, despite the near-ubiquity of the phenomenon, an explanation for the TSR remains elusive, with multiple hypotheses proposed [13-15]. Problematically, the physiological processes that regulate body size with respect to temperature are largely unknown, an important first-step for generating a mechanistic understanding of the TSR.

A number of mechanistic hypotheses have been proposed to explain the TSR [8, 13-19] and each differ in how temperature is thought to affect the developmental and physiological processes that regulate body size. For example, the van der Have and de Jong model of TSR [18] proposes that temperature directly regulates the rate and duration of growth, but the rate of growth is less thermally sensitive than the rate of differentiation, which controls the duration of development and growth. Consequently, as temperature increases, the duration of growth decreases more than the increase in the rate of growth, resulting in an overall reduction in final body size. In contrast, the von Bertalanffy/Perrin model of TSR proposes that temperature directly regulates final body size and growth rate, but not growth duration [20, 21]. Under this hypothesis, the cessation of growth occurs at a size where the rate of anabolism (energy acquisition) balances the rate of catabolism (energy loss). If an increase in temperature enhances catabolism more than anabolism, then this balance will be achieved at a smaller size at higher temperatures. At the same time an increase in temperature increases growth rate. The result is that at elevated temperatures, growth duration is reduced because animals grow more quickly to a smaller final size.

Under both the van der Have and de Jong model and the von Bertalanffy/Perrin model, the observed effect of temperature on final size, growth rate and duration is the same. What is different is the proposed locus of regulatory control. Nevertheless, both models are based on ostensibly general thermodynamic features of specific cellular and physiological processes. Implicit in these models are two assumptions. First, thermal plasticity of body size is a result of unavoidable effects of temperature on growth and differentiation, that is, the response of body size to temperature represents a biophysical constraint [*sensu* 14-16, 18]. Second, the thermal reaction norm of size at the whole organismal level can be entirely explained by the thermal sensitivity of processes that occur at the level of cellular biochemistry.

Alternative views suggest the TSR is an adaptation and does not represent a biophysical constraint [8, 13-16]. Natural populations show the same trend as TSR: that is, individuals from populations found in colder climates are typically larger than individuals from populations found in warmer environments, a phenomenon commonly known as the Bergmann's rule [22]. This is not simply a phenotypically plastic response, as the size difference is maintained even when individuals from different geographic populations are reared at the same developmental temperature [23, 24]. Bergmann's rule has been observed repeatedly in both endotherms (for which it was originally formulated) and ectotherms, while laboratory selection experiments have also demonstrated that populations evolved in colder temperatures have larger body size than those evolved in warmer temperatures [25-27]. Although the selective advantage of a small body size at higher temperatures remains unclear, these observations do suggest that the TSR is an adaptive response. If it is an adaptation, the mechanistic regulation of the TSR need not be the same in different taxa, in contrast to a biophysical constraint that is expected to be uniform across all organisms [13-16]. Assuming that, in ectotherms, growth rate increases with temperature, a reduction in body size can then be mechanistically achieved either by reducing the duration of growth, or by triggering the cessation of growth at a smaller body size.

A resolution to the TSR debate may be achieved through a better understanding of the proximate mechanisms of body size regulation and how these mechanisms are influenced by temperature. However, there has been surprisingly little research to elucidate the developmental mechanisms that regulate body size with respect to temperature, and even less work linking these mechanisms to adaptive theories of the TSR [28]. The mechanistic details of size regulation have perhaps been best elucidated in the holometabolous insects, such as the tobacco hornworm *Manduca sexta* and the fruitfly *Drosophila melanogaster* [4-6, 29]. Holometabolous insects grow through a series of larval molts before metamorphosing into their final adult form. Because growth of the adult is constrained by a stiff exoskeleton, final adult size is regulated by the size at which the larva stops growing and initiates metamorphosis. Although the cessation of growth is caused by an increase in the level of circulating ecdysteroids, the decision to metamorphose is made much earlier in the final instar, at the attainment of a critical size [5, 30-32]. Attainment of critical size is associated with the initiation of a hormonal cascade that ultimately leads to an increase in the level of circulating ecdysteroids, which causes the cessation of growth and the beginning of metamorphosis. Maximum larval size in holometabolous insects is therefore regulated by the critical size, plus the amount of growth that is achieved between attainment of critical size and the cessation of growth [29, 33]. This period is called the terminal growth period [TGP] in *Drosophila* [33, 34] and the interval to cessation of growth [ICG] in *Manduca* [4, 29]. Temperature could therefore affect final body size by influencing the critical size, the duration of the TGP/ICG and/or the rate of growth during the TGP/ICG. Further, although peak larval mass is regulated by critical size and growth during the TGP/ICG, there is considerable mass loss during metamorphosis [35, 36], providing an additional mechanism by which temperature could affect final body size.

The observation that holometabolous insects initiate the cessation of growth well before they actually stop growing suggests that the von Bertalanffy/Perrin [21] model cannot explain the TSR in these animals. Evidence from *Manduca* does, however, support the van der Have and de Jong model of the TSR [31, 37]. An increase in temperature increases growth rate but more substantially decreases the ICG, resulting in a reduction in peak larval mass and hence a smaller final body size. Critical size is not affected by temperature in these insects. Nevertheless, whilst thermal plasticity of body size appears to be mediated by the differential effects of temperature on growth rate and duration, this need not be because of underlying biophysical constraints. Rather it could be a result of selection targeting the mechanisms that regulate the duration of the ICG and/or growth rate. Importantly, if the adaptive hypothesis were correct, then we might expect selection to target the thermal sensitivity of other size regulatory mechanisms in other holometabolous insects, for example critical size or mass loss during metamorphosis.

Here we test the hypothesis that the TSR is regulated by different developmental mechanisms in different insects, by identifying these mechanisms in a second holometabolous insect, *Drosophila melanogaster*. Our data indicate that unlike *Manduca sexta*, temperature influences final body size primarily by regulating critical size in *Drosophila*, that is the size at which larvae initiate the cessation of growth. Collectively, our data do not support a common mechanism to explain the TSR but suggest that the TSR reflects alternative developmental responses to the same selective pressure.

## Methods

*Fly stocks and rearing*

Flies were from an isogenic stock of the Samarkand strain (SAM). For the measurement of larval growth trajectories at 25°C female larvae also carried a ubi-GFP transgene (Bloomington Stock Center, 1681) that had been backcrossed into SAM for five generations. This allowed us to conduct experiments exploring the developmental regulation of sexual size dimorphism, described elsewhere (Testa, Ghosh, Shingleton, *unpublished data*). All experiments were conducted at constant-light regime and fly cultures were maintained at low density (50-60 larvae per 6 mL food) on standard cornmeal-molasses medium.

*Measurement of critical size*

Critical size was measured following the method used in Stieper et al. 2008 [38]. Female flies were allowed to oviposit on fresh plates (60×15 mm Petri dishes filled with 10 ml of standard cornmeal/ molasses medium) at 25°C for 24 h. Eight such plates were incubated to 17°C, and the remaining eight were kept at 25°C. Each plate contained approximately 100 eggs. For 25°C, 3 days after the endpoint of egg lay (AEL), individual third instar larvae (L3) were withdrawn from food, weighed using Metler Toledo XP26 microbalance (d= 0.001 mg), and placed in a 1.5ml microtube with a 10 x 50 mm strip of moist KimWipe. Larvae were inspected for pupariation every 4h and re-weighed if pupariated. The procedure was repeated for larvae reared at 17°C, 6 days AEL.

*Calculating critical size*

Critical size was calculated using the breakpoint method detailed in Stieper et al. 2008 [38] with some modifications. Third instar larvae smaller than critical size show a delay in the initiation of the hormonal cascade that ends in metamorphosis, compared to fed larvae of the same size [39]. In contrast, starvation after critical size actually accelerates metamorphosis relative to larvae that are allowed to continue feeding. Because of this change in response to starvation, a plot of larval weight at starvation against time to pupariation (TTP) shows a significant change in slope, or a breakpoint, at critical size (figure 3), which can be detected using a bi-segmental linear regression. The performance of the bi-segmental linear regression is improved if one plots the corresponding pupal weight for a starved larva against TTP, and further improved if the data is rotated 5 rad around the origin prior to analysis. We therefore (i) plotted the relationship between pupal weight and TTP, (ii) rotated the plot by 5 rad around the origin, (iii) detected the pupal weight and TTP at the breakpoint using the *segmented* package in R [40], (iv) back-rotated these values by -5 rad, and (v) converted the pupal weight to a larval weight using the parameters of a linear regression of pupal weight against larval weight at starvation (here referred to as the converted larval weight) for each temperature. We repeated the analysis on 1000 bootstrap data sets to generate 95% CI for the critical size and TTP at critical size (TTP$_{CS}$). We used a permutation test with 1000 replicates to generate a null distribution of the difference in critical size and TTP$_{CS}$ at the two temperatures, and used this distribution to estimate a *P*-value for the observed differences.

*Measurement of growth curves*

Eggs were laid on 12 food plates for 4h at 25°C. Six plates were kept at 25°C and the remaining six were incubated at 17°C. Approximately 35 larvae were randomly sampled from the 25°C plates at 12h and 24h after hatching (AH), and subsequently every 6h until 84h AH. Larvae were washed in distilled water, dried on a KimWipe and weighed. The same approach was used to measure growth rate at 17°C, except larvae were weighed every 8h from 8h AH to 200h AH. At 25°C, the first pupariations were observed at 84h AH and at 90h, the plates were cleared of all pupae. A cohort of 24 pupae formed between 90 to 94h AH was subsequently collected, cleaned and weighed, and each pupa was placed on a moist piece of Kimwipe and kept inside a microtube. The pupae were weighed every 24h until eclosion, and the wet weight of the emerging adults were recorded. At 17°C, the plates were cleared of pupae at 212h, and 40 pupae were collected at 216h AH; pupal and adult weights were measured following the same methods as 25°C.

*Calculation of growth parameters*

We used the mean mass of larvae/pupae within each age-cohort to generate growth curves at 17°C and 25°C and to calculate key growth parameters. We define the TGP as the interval between attainment of critical size and the age at which the larva ceases to grow further, which coincides with the start of wandering behavior. We used our larval growth curves and calculation of the critical size to determine the age at which larvae achieve critical size. We calculated the age at which the cessation of growth occurs by using multiple comparisons (Hsu's MCB test) to identify the age-cohort at which mean larval mass was not significantly different from the age-cohort with the greatest mean mass ($P$>0.05). Subtracting the age at which larvae attain critical size from the age at growth cessation gave the duration of the TGP. Subtracting critical size from the individual larval masses within the age-cohort at growth cessation gave the amount of mass gained during the TGP, which was compared between temperatures using a t-test. Logarithmic growth rate during the TGP was calculated by regressing (log) larval mass against age and comparing between temperatures using an ANCOVA. Mass lost during larval wandering and metamorphosis was calculated by comparing the masses of individuals in the age-cohort at the cessation of larval growth with the masses of individuals in the age-cohort at the end of pupation, which was then compared between temperatures using an ANOVA.

We used nominal logistic regression to predict the age and mass at which 50% of larvae have transitioned to the next developmental stage at each temperature, allowing us to match growth with development. Finally, we measured $3^{rd}$ instar mouth-hooks in 10 larvae reared at 17°C and 25°C: larval mouth-hooks reflect the larval size at the previous molt and larval growth achieved through the preceding instar [41].

*Statistical analysis*

Statistical analyses were conducted using R (www.r-project.Org) and JMP (SAS Institute). Data were tested to confirm normality of error, linearity and homogeneity of variance where necessary [42].

## Results

As with other insects, *Drosophila* larvae grow significantly larger when reared at lower temperatures. Newly eclosed adults weigh 1.00 mg (95% CI, 0.92-1.08mg) when reared at 17°C but only 0.88 mg (95% CI, 0.80-0.96 mg) when reared at 25°C (figure 1, figure 2F) (t-test, *P*<0.05). This difference in final body size is reflected in a significant difference in critical size, which is 1.10 mg at 17°C (95% CI: 1.03-1.17 mg) and 0.87 mg at 25°C (95% CI: 0.83-0.9 mg) (permutation test, *P*<0.001) (figure 2A).

The duration of each larval instar is, as expected, significantly longer at 17°C than at 25°C (figure 1) (nominal logistic regression, *P*<0.001 for timing of $1^{st}$-$2^{nd}$ instar ecdysis and $2^{nd}$-$3^{rd}$ instar ecdysis). However, the mass of larvae at the end of the first- and second-larval instar is not significantly different at the two temperatures (nominal logistic regression, *P*=1.00 for $1^{st}$-$2^{nd}$ instar molt, *P*=0.21 $2^{nd}$-$3^{rd}$ instar molt). This suggests that the size difference between larvae reared at the two temperatures do not appear until they reach the third and final instar. Consistent with this, we found that the third instar mouth-hook size, which indicates the size of the larva at the previous molt, is not significantly different between temperatures (t-test, $P$=0.2718).

The shapes of the growth curves appear ostensibly similar at 25°C and 17°C as depicted in figure 1, except that the duration of development is much longer at 17°C. Larvae grow significantly more rapidly at 25°C than 17°C, with logarithmic growth rates during the TGP of 0.037mg.h$^{-1}$ and 0.013mg.h$^{-1}$, respectively (ANCOVA, $P$=0.0148) (figure 2B). Higher temperature, however, greatly shortens the duration of TGP from 48 h at 17°C to 18 h at 25°C (figure 1,2C). Our data do not allow us to statistically compare the length of TGP at the two temperatures. As an alternative we can use the time to pupariation from critical size (TTP$_{CS}$), which we measured in our critical size assay, as a proxy for TGP. TTP$_{CS}$ is significantly shorter at 25°C compared to 17°C (permutation test, $p < 0.001$) (figure 2C). Interestingly, the net effect of temperature on the growth rate during the TGP and the duration of the TGP are such that the mass gained during the TGP is not significantly different at the two temperatures (figure 2D) (t-test, $P = 0.098$).

After reaching peak larval mass, larvae start wandering and lose mass during pupariation and metamorphosis. The average mass loss between peak larval mass and adult eclosion is 0.8 mg at 25°C (95% CI: 0.68-0.92 mg), and 0.83 mg at 17°C (95% CI: 0.72-0.95 mg), which is not significantly different (figure 2E) (ANOVA, $P = 0.70$).

Not only are the mass gained post critical size and subsequent mass loss the same across the two treatment temperatures 25°C and 17°C, but the two processes largely cancel each other for any given temperature. As a result, the thermal plasticity of adult weight largely reflects the thermal differences in critical size (figure 2A,F).

The effect of temperature on critical size is clearly evident from the plots of larval mass at starvation against time to pupariation: the breakpoint in the plots is at a greater mass for 17°C larvae compared to 25°C larvae (figure 3). This breakpoint indicates a change in the relationship between larval mass at starvation and the subsequent time to pupariation, which arises because starvation before attainment of critical weight delays the initiation of the hormonal changes that cause pupariation, while starvation after critical weight does not. The slopes of the relationship before and after the breakpoint reflect, in part, larval growth rate. It is possible that the difference in the position of the breakpoint at 17°C and 25°C is largely a consequence of differences in the slopes either side of the breakpoint, which in turn reflects the differences in larval growth rate at the two temperatures. To confirm this was not the case, we re-analyzed our data using time in degree-days (DD) rather than hours (see electronic supplementary material for details). This has the effect of equalizing larval logarithmic growth rate and $TTP_{CS}$ at the two temperatures (electronic supplementary figure 1A,B). Nevertheless, the critical sizes is unchanged and remains significantly different at the two temperatures: 1.07 mg at 17°C (95% CI: 0.96-1.13 mg) and 0.86 mg at 25°C (95% CI: 0.81-0.90 mg) (permutation test, *P*<0.001) (electronic supplementary figure 1C,D).

To confirm that the difference in critical size across temperatures is biologically accurate rather than a consequence of our method of analysis, we also calculated the minimal viable weight for pupariation, MVW[P], which has been used as a proxy for critical size in *Drosophila* [43-45]. MVW[P] is defined as the minimal weight at which 50% of larvae survive to pupariation when starved [41, 44], and is attained approximately the same time as critical size in *Drosophila* [32, 41]. We used a nominal logistic regression to predict the weight at which the probability that a starved larva survives to pupariation is 0.5. The MVW[P] is 1.10 mg at 17°C (95% CI: 1.05-1.15 mg) and 0.84 mg at 25°C (95% CI: 0.80-0.89 mg), thus supporting our hypothesis that critical size in *Drosophila* indeed changes with temperature.

## Discussion

Our data indicate a novel mechanism for the regulation of the temperature-size rule in *D. melanogaster*; that is, the TSR results from the thermosensitivity of the critical size. Critical size is a 'decision point' when larvae commit to metamorphosis and its attainment initiates an endocrine cascade that eventually results in the cessation of growth [30]. Reduced critical size at a higher temperature therefore suggests that larvae instigate the signal to stop growth at a smaller size. Although larvae accumulate additional mass between critical size and the cessation of growth, followed by mass loss during wandering and metamorphosis, the change of mass during these phases does not vary across temperature. As a result, smaller adult size at a higher temperature in flies arises solely from the thermal reduction in critical size.

It is possible that the thermal regulation of adult weight by critical size may be specific for the genotype of flies we used (Samarkand strain, SAM), and the thermal range used in the study (17°C and 25°C). However, published and unpublished data from our laboratory show that critical size for SAM and a second wild-type strain, OreR, is the same at 25°C and significantly lower at 29°C in OreR [37, Shingleton *unpublished data)*. Thus critical size appears to be the locus of thermal regulation of adult size in flies of different genotypes across a broad thermal range.

*Regulation of critical size*

Our data indicate that the critical size is temperature sensitive in *Drosophila* and that it underlies the developmental basis of the TSR. It is, however, unclear how this sensitivity is achieved mechanistically. Work over the last decade has indicated that critical size is regulated by at least two signaling pathways; the insulin/IGF-signaling pathway and the PTTH/Ras/Raf signaling pathway [41, 45, 46-49]. Both appear to control the timing of ecdysteroid synthesis by the prothoraric gland and the cessation of growth: insulin/IGF signaling apparently in response to nutritional status [41, 46, 47]; PTTH/Ras/Raf signaling in response to temporal information [45, 48]; and both insulin/IGF and PTTH/Ras/Raf signaling in response to the developmental status of the growing organs [32, 50-52].

However, it is unclear whether these or some other pathway mediates the effects of temperature on critical size.

Recently, an additional regulator of critical size has been proposed – oxygen level [53]. In all holometabolous insects, tracheal volume is largely fixed within an instar and only expands at each larval molt. Consequently, as a larva grows its tracheal system becomes limiting for oxygen delivery, and in *Manduca sexta* this is evident as a leveling off of whole-animal respiration rates midway through each larval instar. Intriguingly, in the final instar this plateau is reached at critical size, suggesting that *Manduca* larvae sense their size and attainment of critical size through a decline in oxygen availability. This is supported by evidence that critical size is reduced in *Manduca sexta* larva reared in low oxygen conditions. Because an increase in temperature increases metabolic rate [54], the size at which oxygen becomes limiting should be lower at higher temperatures, assuming tracheal volume does not change. Although we do not know the effect of temperature on tracheal volume in *Drosophila* larvae, our finding that larvae reared at 17°C and 25°C are the same size at ecdysis to the third instar suggest that tracheal volume is unaffected by temperature. An intriguing hypothesis, therefore, is that *Drosophila* larvae exploit the effects of temperature on the timing of oxygen limitation as a mechanism to adjust their critical size, and hence body size, with temperature. Effect of oxygen level on adult size in *Drosophila* is more pronounced at higher temperatures supporting the view that oxygen supply may play a role in thermal regulation of size in flies [54].

*Diverse modes of achieving the TSR*

Our observations of the thermal regulation of body size in flies are in contrast to previous findings in another holometabolous insects, the tobacco hornworm *M. sexta* [4, 31, 37]. As discussed above, in *Manduca sexta* critical size does not change with temperature [31]. Rather, the effect of temperature on final body size is mediated by the amount of growth achieved during the TGP/ICG, which is reduced at higher temperatures because the increased growth rate cannot compensate for the shortened TGP/ICG [4, 29]. Thus the TSR appears to be achieved using different mechanisms in the two species.

It is important to note, however, that in contrast to our experiment, the effects of temperature on the mechanisms that regulate body size in *M. sexta* were assayed in larvae reared at the same temperature until the final instar, and then moved to different temperatures for the rest of development [31]. It is possible, therefore, that the thermal plasticity of *M. sexta* is also regulated solely by critical size in larvae reared at a single temperature for the duration of development. This seems unlikely. In *M. sexta*, early developmental temperature influences the size of larvae entering their final instar [55], which we did not observe in *Drosophila*. Further, in *M. sexta* larvae reared at a constant higher temperature, there is still a lack of compensation between a decreased growth period and an increased growth rate during the final instar [55], which we did not observe in *Drosophila*.

*Implications for the evolution of the TSR*

A central issue in explaining the temperature size rule and its ubiquity is distinguishing between proximate (mechanistic) and ultimate (evolutionary) causes. Non-adaptive explanations for the TSR explicitly deny an evolutionary explanation and propose that thermal plasticity in body size is a result of biophysical constraints imposed by temperature on developmental and physiological processes. The thermal sensitivities of anabolism versus catabolism [20, 21], and developmental rate versus growth rate [18, 19] have both been proposed as general non-adaptive explanations of the proximate TSR mechanisms. However, the observation that natural and experimentally evolved populations show the same trend as the TSR argues strongly that the TSR is an evolved response to selection for a reduced body size at higher temperatures. Further, many of the details of the biophysical models do not appear to be sufficiently general to act as proximate regulators of body size in animals with diverse modes of development [14, 15, 56, 57]. Finally, non-adaptive biophysical explanations for the TSR imply that the effects of temperature on body size are mediated at the level of cellular and sub-cellular processes [18, 27, 58, 59]. However, our data indicate that, in *Drosophila* at least, the TSR is achieved through a physiological mechanism (attainment of critical size) that acts at the level of the whole organism.

If the TSR is an adaptation, what developmental mechanisms have been targeted by selection to generate the appropriate thermal plasticity in body size? In ectotherms there is a general trend that growth rate increases with temperature and this is thought to be due to biophysical kinetics of the enzymatic reactions that regulate metabolism [60]. If there is selection for ectotherms to be smaller at higher temperatures and they are constrained to grow faster at higher temperatures, then it follows that there will be selection to reduce their duration of growth at higher temperatures. Crucially, this reduction in growth duration must be greater than the coincidental increase in growth rate. There are two ways an animal can reduce the duration of growth at higher temperatures. Either the molecular and physiological processes that control development can progress at a faster rate, or an animal can stop growing at a smaller size. Problematically, at the level of whole-animal growth trajectories these two mechanisms appear identical and can only be distinguished through elucidation of the mechanisms that regulate the rate and duration of development.

Our data suggest that *D. melanogaster* and *M. sexta* utilize different mechanisms to regulate the duration of growth at different temperatures and control thermal plasticity of body size. In *M. sexta,* elevated temperature reduces the duration of the TGP/ICG more than it increases growth rate, so that while critical size is temperature insensitive, the amount of mass gained during the TGP/ICG is reduced at higher temperatures. In *Drosophila,* the amount of mass gained during the TGP/ICG is temperature insensitive, but the critical size is reduced at higher temperatures. Consequently, in both groups, the total growth duration is more sensitive to temperature than growth rate, but in the latter this is because larvae initiate the cessation of growth at a smaller size at higher temperatures. The effect of temperature on critical size in *D. melanogaster* larvae is unlikely to be a consequence of development progressing at a higher rate relative to the rate of growth at higher temperatures. This is because larvae molt to the second and third larval instars at the same size at both 17°C and 25°C. Rather, we predict that the thermal plasticity of critical size reflects the effect of temperature on the mechanisms fly larvae use to assess their size.

While our data suggest that *D. melanogaster* and *M. sexta* have evolved to obey the TSR using different mechanisms, it is not possible to say which of the physiological processes that control body size (critical size, growth rate, duration of TGP, and weight loss post TGP) were the proximate target of selection. Temperature affects myriad physiological processes in developing ectotherms, and just because a physiological process is thermosensitive does not mean that this thermosensitivity is an adaptation. Selection for the TSR in holometabolous insects may modify the thermosensitivity of one process, for example growth rate, to accommodate the response of another process to temperature, for example the duration of the TGP. It is possible, therefore, that thermal sensitivity of critical size in *Drosophila* is not an adaptation *sensu stricto* but a biophysical consequence of how temperature affects the mechanisms larvae use to assess their size. Under this hypothesis, selection may have targeted thermosensitivity of the processes that regulate mass gain during the TGP to ensure that body size correctly matches developmental temperature. Nevertheless, the observation that critical size is temperature sensitive in *Drosophila* but temperature insensitive in *Manduca* suggests that selection has targeted the response of critical size to temperature in at least one of these species, although it does not tell us which one.

Our findings support the hypothesis that the TSR is adaptive, but do not address the question of what it is an adaptation to. There are several adaptive models to explain the ubiquity of the TSR [14, 61]. In general, these adaptive models take a 'top-down' approach, starting by hypothesizing the selective pressures that generate the TSR. The selective forces include: optimizing resource allocation between growth and reproduction [62]: optimizing the timing of maturation [61, see discussion in 14]: optimizing the oxygen delivery in case of aquatic species [63]; or maximizing starvation resistance [64]. Nevertheless, a universal adaptive explanation remains elusive. Indeed, there are arguments against the concept that there is a unifying explanation for the existence of the TSR across taxa [15, 65]. An alternative approach is to start by elucidating the proximate developmental mechanisms that regulate the TSR, which will allow elucidation of the focal traits on which selection acts to generate the TSR. This in turn facilitates identification of what these underlying selective pressures are. Consequently, we propose that a 'bottom-up' approach is likely to provide a fruitful and complementary approach to understand the adaptive nature of the TSR.

*Conclusion*

Our findings suggest *Drosophila melanogaster* uses a novel mechanism for the regulation of the TSR. In flies, individuals reared at higher temperature initiate the signals to stop growing at a smaller size, which is different from the mechanism used to generate the TSR in the moth, *Manduca sexta*. This suggests selection for the TSR targets different developmental mechanisms in different taxa. Because *Drosophila* is such a tractable model for studying the molecular-genetic basis of development, this study is foundational to elucidating the proximate details of the TSR. This elucidation is essential for the complete understanding of the adaptive significance of the TSR.

## Acknowledgements

We are grateful to Viviane Callier for comments on early drafts of this manuscript. We thank Austin Dreyer, Rewatee Gokhale, Nathan Parker and Johnathan Constan for assisting in data collection. This work was supported by the National Science Foundation (Grant No. IOS-0845847).

## Author Contributions

Conceived and designed the experiments: SMG AWS. Performed the experiments: SMG NDT. Analyzed the data: SMG NDT AWS. Contributed reagents/materials/analysis tools: AWS. Wrote the paper: SMG AWS.

## Figure legends

Figure 1: Complete growth curves from larval hatching to adult eclosion at 25°C (black curve) and 17°C (gray curve). Circles are larval and pupal masses and squares are final adult masses. The difference between final pupal mass and adult mass is due to loss of pupal case. Error bars are standard errors and are obscured by the markers in some cases. Upper panels show duration of each developmental stage at the two temperatures, with vertical lines indicating when 50% of larvae have made the developmental transtion (L1 = $1^{st}$ larval instar, L2 = $2^{nd}$ larval instar, L3 = $3^{rd}$ larval instar).

Figure 2: Effect of temperature on key size-regulating developmental parameters. Black bars depict values for 25°C, grey bars depict 17°C. (A) Critical size calculated using the breakpoint method (B) Logarithmic growth rate during TGP. (C) Duration of TGP and $TTP_{CS}$. (D) Mass gained during TGP. (E) Mass loss

during pupariation and metamorphosis. (F) Wet weight at eclosion. All error bars are 95% confidence intervals. See text for details of statistical comparisons.

Figure 3: Time to pupariation as a function of converted larval weight at starvation. The broken lines show the breakpoint in the data and indicate the critical size at 25°C (black) and 17°C (gray and black).

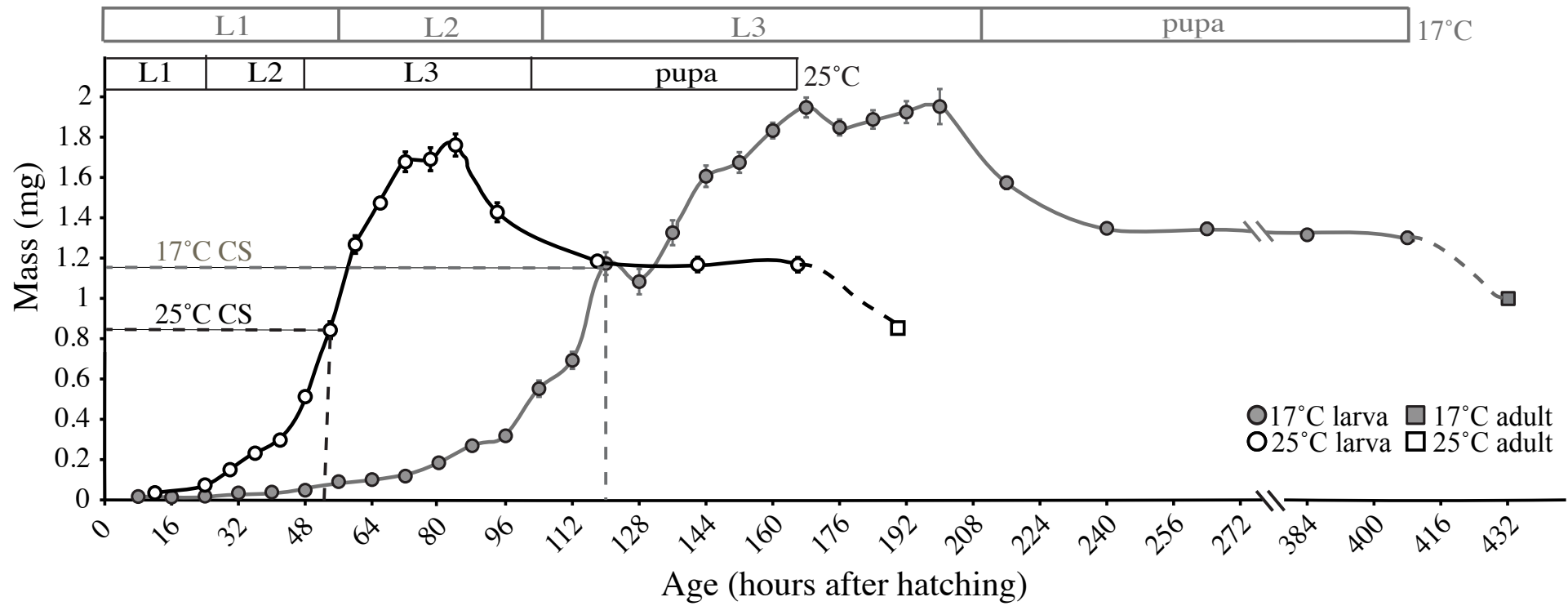

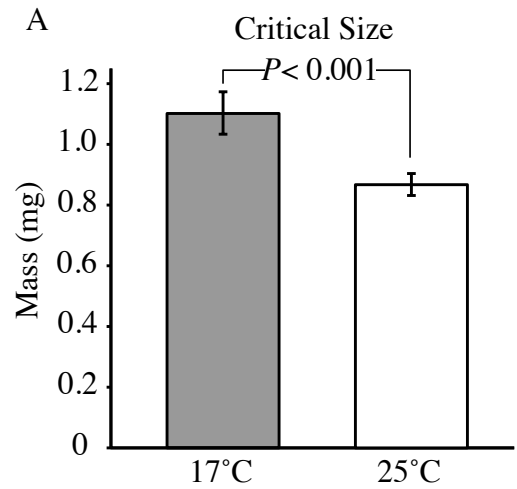
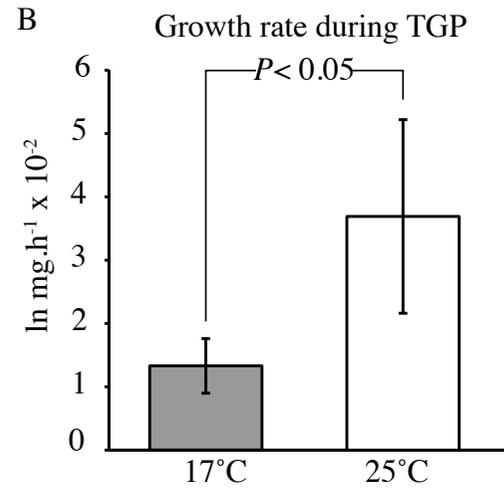
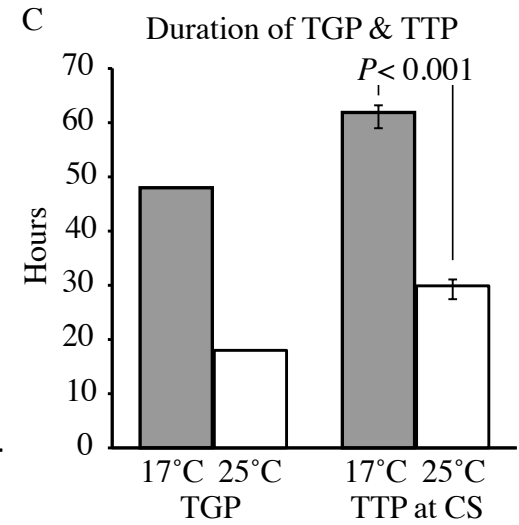
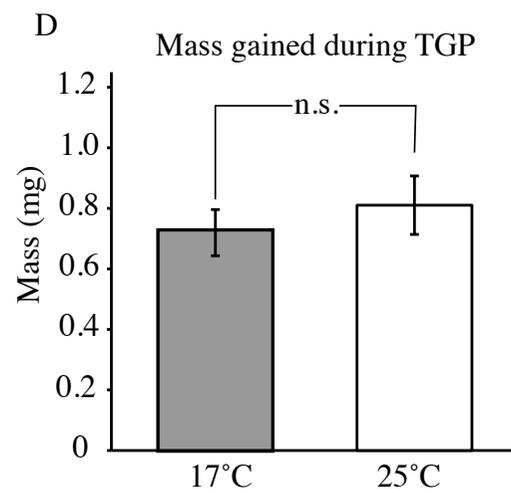
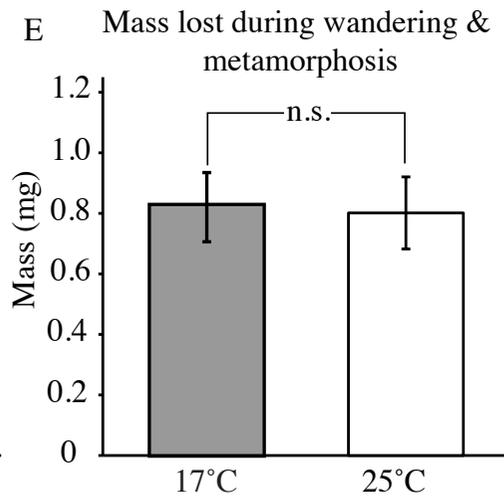
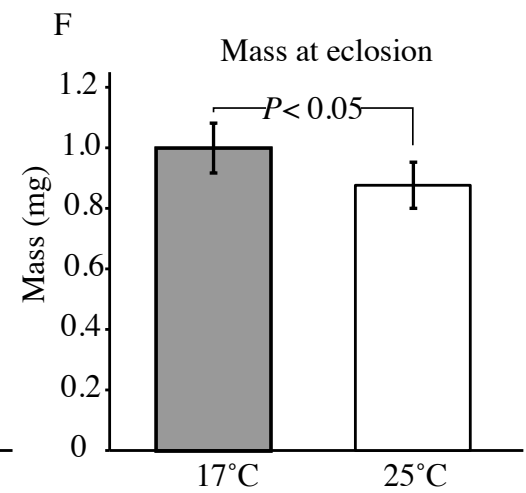

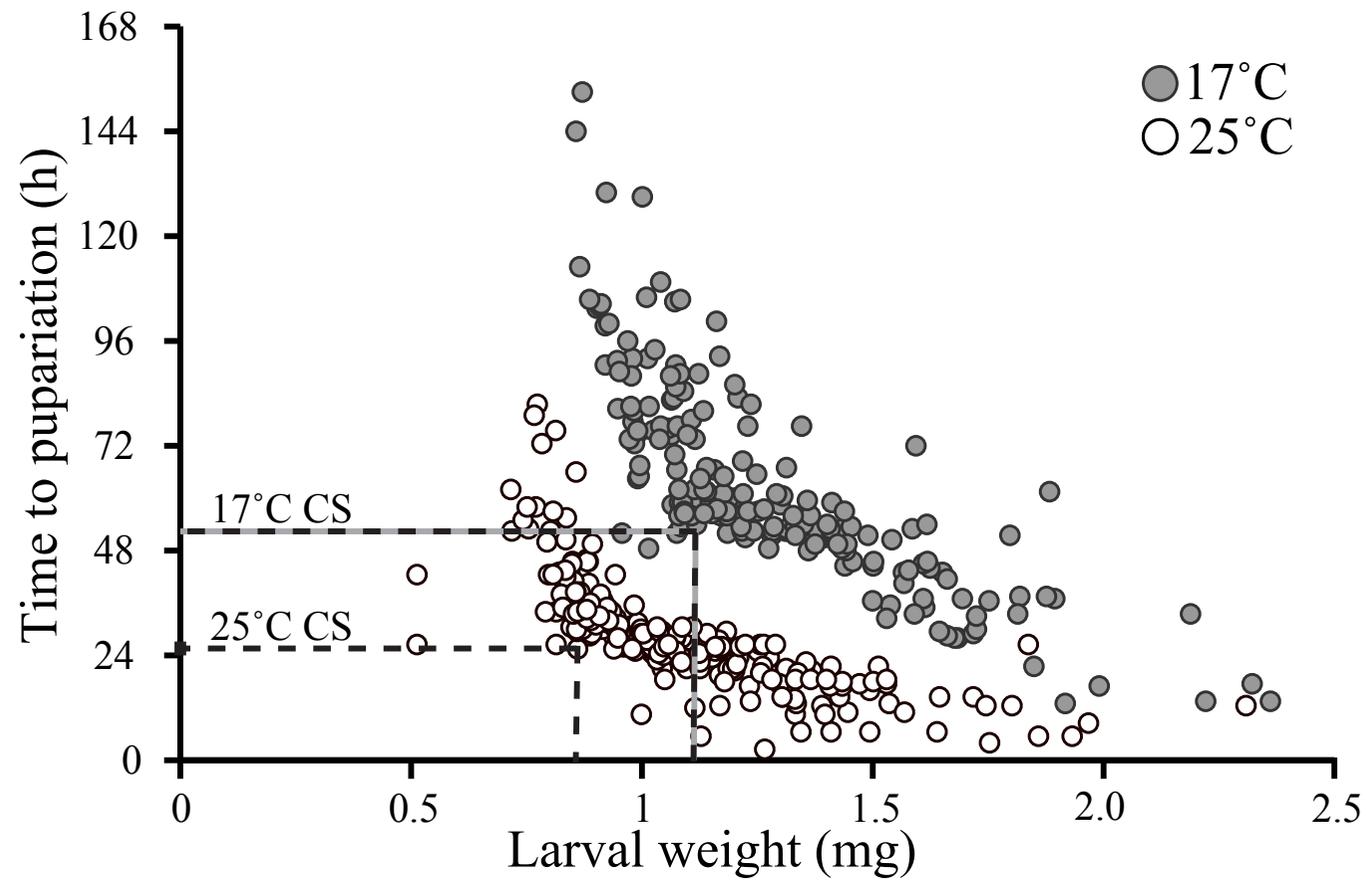

**Electronic Supplementary Material:**

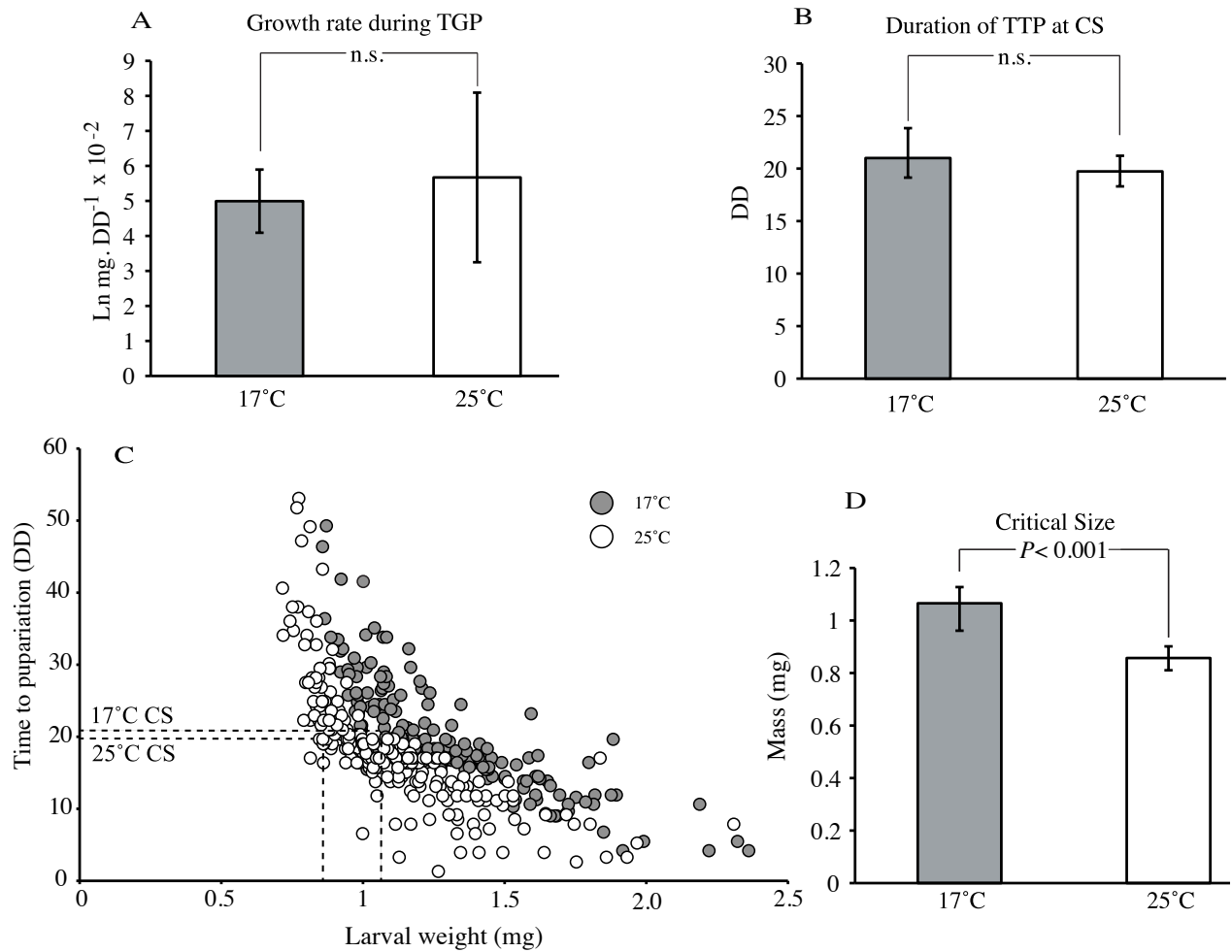

Figure 1: Effect of temperature on key size-regulating developmental parameters calculated using degree-days (DDs). DD is a measure of heat accumulation above a minimal temperature ($T_0$), and was calculated as $(T_r - T_0)n_r$, where $T_r$ is the developmental temperature and $n_r$ is time (days) maintained at $T_r$. We calculated the value of $T_0$ (9.27°C) such that the total developmental development time in DD from hatching to adult eclosion was the same at 17°C and 25°C (136.8 DD)[1]. All measures of larval age and TTP at starvation where then converted into DD. Black bars depict values for 25°C, grey bars depict 17°C. (A) Logarithmic growth rate in DD during TGP. (B) Duration of $TTP_{CS}$ in DD. (C) Time to pupariation in DD plotted against converted larval weight at starvation. The broken lines show the breakpoint in the data and indicate the critical size at 25°C (black) and 17°C (gray and black). (D) Critical size calculated using the breakpoint method. All error bars are 95% confidence intervals.